\documentclass[american]{article}
\usepackage[T1]{fontenc}
\usepackage[utf8]{inputenc}
\usepackage{babel}
\usepackage{url}
\usepackage{amsmath}
\usepackage{amssymb}
\usepackage{graphicx}
\usepackage{esint}
\usepackage[bookmarks=false,
 breaklinks=false,pdfborder={0 0 1},backref=false,colorlinks=false]
 {hyperref}

\makeatletter
\usepackage{babel}
\makeatother

\begin{document}
\title{Strong Likelihood Principle: Strengthening a Principle or Misunderstanding
the Likelihood Function}
\author{Paul William Vos}
\maketitle
\begin{abstract}
The strong likelihood principle (SLP) is conventionally derived from
the sufficiency principle and a conditionality principle in an argument
due to Birnbaum, and much of the literature contests whether the derivation
is sound. We take a different approach. We ask what the SLP says when
its terms are read carefully, and argue that the principle as ordinarily
stated reflects a confusion about the domain of the likelihood function.
The likelihood is naturally defined as a function on a family of distributions
$M$, not on a parameter space, and once it is so defined the SLP
collapses into its weak counterpart, the weak likelihood principle.
The diagnosis is illustrated by analogy with monetary value, developed
concretely through a comparison of the binomial and negative binomial
families that share a parameter, and connected to the geometric structure
of $M$ through the Fisher information metric. The same standardization
emerges from a statistical argument about comparing measurements across
populations and from a geometric argument about manifold distance;
this convergence supplies the positive content of the weak likelihood
principle. 
\end{abstract}

\section{Background}

The strong likelihood principle (SLP) asserts that two outcomes producing
proportional likelihood functions carry the same evidential content
for inference about a shared parameter, even when the outcomes arise
from experiments with different probability models. \cite{birnbaum1962}
argued that the SLP is a logical consequence of two more elementary
principles, sufficiency (SP) and conditionality. The argument has
the form of a conditional, with SP and a version of conditionality
as antecedents and the SLP as consequent, and the substantial literature
that has grown up around it can be sorted by where critical attention
has been directed.

Most of that attention has been directed at the antecedents. The exchange
between \cite{mayo2014} and \cite{dawid2014} is representative.
Mayo, working with Cox's weak conditionality principle (WCP), constructs
sampling-theoretic outcomes that violate the SLP while satisfying
both SP and WCP, and concludes that the Birnbaum derivation fails.
Dawid replies that Birnbaum's original conditionality principle is
not the WCP; Birnbaum's principle is a non-directional equivalence,
while the WCP is a directional injunction to condition on the experiment
actually performed. The principles being compared, Dawid maintains,
are not the same, and Birnbaum's theorem stands. Earlier criticisms
in the same vein---\cite{fraser1963}, \cite{durbin1970}, \cite{kalbfleisch1975},
\cite{evansFraserMonette1986}---propose various modifications of
SP or conditionality that likewise block the derivation. The common
feature of these treatments is that they take the consequent of the
argument, the SLP, as the principle whose content is settled, and
contest the meaning or applicability of the principles offered as
antecedents.

The present paper takes the opposite tack. We set aside the question
of whether the antecedents, however formulated, entail the consequent,
and ask what the consequent itself says. Our claim is that the SLP,
as ordinarily stated, presupposes an understanding of the likelihood
that does not survive careful examination: namely, that the likelihood
is a function of a parameter rather than a function of a distribution.
Once the likelihood is treated as a function on the family of distributions
$M$, with parameter values serving as labels for elements of $M$,
the SLP collapses into its weak counterpart, the weak likelihood principle
(WLP), and the argument over Birnbaum's proof becomes an argument
over a principle that no longer says anything distinctive.

The remainder of the paper develops this position. Section \ref{sec:parameter-free}
gives a parameter-free definition of the likelihood and shows that
the WLP and SLP coincide once the domain is correctly identified.
Section \ref{sec:dollar-principle} illustrates the same point through
an analogy with monetary value. Section \ref{sec:role-of-parameters}
introduces parametric expressions of the likelihood, discusses the
role of parameters as units of measurement, and examines how the score
function and Fisher information acquire their geometric meaning as
objects on $M$ rather than on $\Theta$. Section \ref{sec:score-principle}
considers the score principle implied by the SLP, develops the parallel
statistical practice of standardization for comparing measurements
across populations, and observes that the geometric standardization
of the previous section and the statistical standardization of the
score coincide.

\section{Parameter-free Definition}

\label{sec:parameter-free}

This section gives a definition of the likelihood that makes no reference
to a parameter. We begin by recording, with a little care, what the
mathematical objects are: a distribution, a family of distributions,
and the likelihood as a function from the family to the positive reals.
The care matters because much of the confusion surrounding the SLP,
in our view, comes from compressing these objects into a single expression
$L(\theta;y)$ that obscures which arguments are varying over what.
Once the objects are distinguished, the statement of the likelihood
principle, and the collapse of the SLP into the WLP, follow with little
additional work.

\subsection*{Distribution $m$ and family $M$}

Let $\mathcal{Y}$ be a sample space carrying a $\sigma$-field $\mathcal{B}$
and a $\sigma$-finite measure $\mu$ on $\mathcal{B}$. A \emph{distribution}
on $\mathcal{Y}$ is a probability measure $m$ on $\mathcal{B}$
that is equivalent to $\mu$. By the Radon--Nikodym theorem each
such $m$ has a density $f_{m}=dm/d\mu$ satisfying 
\[
m(B)=\int_{B}f_{m}(y)\,d\mu(y)\qquad\text{for all }B\in\mathcal{B},
\]
and $f_{m}>0$ a.e.

It will be useful to treat a distribution as a single mathematical
object rather than as a procedure. A function is a set of ordered
pairs, and the symbol $m$ can be identified with either of two such
sets: 
\begin{equation}
m=\bigl\{\bigl(y,f_{m}(y)\bigr):y\in\mathcal{Y}\bigr\},\qquad m=\biggl\{\biggl(B,\int_{B}f_{m}\,d\mu\biggr):B\in\mathcal{B}\biggr\}.\label{eq:m}
\end{equation}
The first is the density, a function on $\mathcal{Y}$; the second
is the measure, a function on $\mathcal{B}$. They contain the same
information---the density determines the measure by integration and
the measure determines the density by Radon--Nikodym---and we use
$m$ for both, context making clear which is meant. When $\mathcal{Y}$
is an open interval of the real line, the reader may find it helpful
to picture $m$ as the corresponding density curve over $\mathcal{Y}$;
the curve is the graph of the function in the first form of (\ref{eq:m}).
The point of writing $m$ as an explicit set of ordered pairs is that
$m$ is then a single object, and a family of distributions 
\[
M=\{m_{1},m_{2},\ldots\}
\]
is a set whose elements are such objects. As a set, $M$ has no further
structure: its elements are points. Most families of interest carry
additional structure---usually that of a smooth manifold---but for
the present purpose $M$ is merely a set.

For a sample $y\in\mathcal{Y}$, the \emph{likelihood} is the function
$L^{y}_{M}:M\rightarrow\mathbb{R}^{+}$ defined by 
\[
L^{y}_{M}(m)=f_{m}(y).
\]
Its domain is $M$. The \emph{log likelihood} $\ell^{y}_{M}:M\rightarrow\mathbb{R}$
is defined by $\ell^{y}_{M}(m)=\log L^{y}_{M}(m)=\log f_{m}(y)$.
There is no parameter in either definition, and none is needed: assessing
how each model in $M$ relates to the observation $y$ requires only
the mass $f_{m}(y)$ each model assigns to $y$. The likelihood principle,
on this reading, is the assertion that for inference within $M$ this
single value attached to each $m$ is all that is required.

The claim is a demanding one. Each $m$ is a rich mathematical object,
and many functions of inferential interest are defined using more
of its structure than the value $f_{m}(y)$: method-of-moments estimators,
for instance, depend on the algebraic structure of $\mathcal{Y}$,
not merely on the masses assigned to its elements. The likelihood
principle directs us to ignore all of this and report a single number
per distribution---the value $f_{m}(y)$, which is the height of
the density at the observed $y$ when $\mathcal{Y}$ is continuous
and the probability assigned to $y$ when $\mathcal{Y}$ is discrete.

\subsection*{Parameter-free quantities}

Many standard objects of likelihood-based inference admit definitions
that make no reference to a parameter: the maximum likelihood estimate,
the Kullback--Leibler divergence, the Hellinger distance, the total
variation distance, and the Bhattacharyya coefficient, among others.
Each is a function of distributions in $M$, not of parameter values.
We give special attention to the maximum likelihood estimate and the
Kullback--Leibler divergence, defined by 
\[
\hat{m}_{y}=\arg\max_{m\in M}f_{m}(y),\qquad KL_{M}(m_{1},m_{2})=\int\log\frac{f_{m_{1}}(y)}{f_{m_{2}}(y)}\,f_{m_{1}}(y)\,d\mu(y).
\]
The familiar parametric expressions are obtained by composing with
the parameterization, e.g.\ $\hat{\theta}_{y}=\theta_{M}(\hat{m}_{y})$,
and depend on the parameterization only in their numerical form, not
in the underlying quantity.

\subsection*{The SLP collapses to the WLP}

The strong likelihood principle asserts that for any two families
$M_{1}$ and $M_{2}$ with respective supports $\mathcal{Y}_{1}$
and $\mathcal{Y}_{2}$, inference from $y_{1}\in\mathcal{Y}_{1}$
and $y_{2}\in\mathcal{Y}_{2}$ should be identical whenever 
\begin{equation}
L^{y_{1}}_{M_{1}}=h(y_{1},y_{2})\,L^{y_{2}}_{M_{2}}\label{eq:SLP-antecedent}
\end{equation}
for some $h(y_{1},y_{2})\in\mathbb{R}^{+}$. The two likelihood functions
in (\ref{eq:SLP-antecedent}) are functions on $M_{1}$ and $M_{2}$
respectively. For one to equal a positive scalar multiple of the other---for
the equality in (\ref{eq:SLP-antecedent}) to hold at all---they
must share a domain. Hence $M_{1}=M_{2}$, and the antecedent of the
SLP reduces to the antecedent of the weak likelihood principle: two
outcomes from the same family with proportional likelihood functions
on that family. The conclusion reduces correspondingly. The SLP, stated
for the likelihood as a function on $M$, is the WLP.

If this is so, the question is why the SLP has appeared to say something
stronger. The answer, anticipated already in the wording of the principle
as \textquotedbl proportional likelihoods\textquotedbl{} with the
constant of proportionality permitted to depend on the data, is that
the likelihood is conventionally written as a function on a parameter
space rather than on $M$, and that two distinct families can share
a parameter space. We turn to parameterizations next. 

A \emph{parameterization} of $M$ is a bijection $\theta_{M}:M\rightarrow\Theta$
onto some index set $\Theta\subseteq\mathbb{R}^{p}$. When $M$ has
further structure (typically that of a smooth manifold), the parameterization
is required to respect that structure---to be a diffeomorphism---but
the choice of diffeomorphism does not matter for expressing the geometry
of $M$. What is essential is that the parameterization be a diffeomorphism.
The parameter itself is never the object of inference; it is a coordinate
system on $M$, in the same way that inches and centimeters are coordinate
systems on a length. We return to this analogy in Section \ref{sec:role-of-parameters}.

The appearance that the SLP says something stronger than the WLP arises
from writing the likelihood as a function on a parameter space and
then comparing across families through their shared parameterization.
Two distinct families $M_{1}$ and $M_{2}$ can be indexed by the
same $\theta\in\Theta$---for example, the binomial and negative
binomial families share the parameter $p\in(0,1)$---and the parametric
likelihoods $L^{y_{1}}_{\Theta}$ and $L^{y_{2}}_{\Theta}$ are then
comparable as functions on $\Theta$ even though the families themselves
differ. The SLP, in its usual statement, asks for inferential equivalence
in this parametric setting. But the equality of $L^{y_{1}}_{\Theta}$
and $L^{y_{2}}_{\Theta}$ up to a constant is not equality of likelihoods;
it is equality of their expressions in a common coordinate. The two
underlying functions $L^{y_{1}}_{M_{1}}$ and $L^{y_{2}}_{M_{2}}$
have different domains and are not the same function. Asserting that
they should generate the same inference is a coordinate-level claim
being made about the objects $M_{1}$ and $M_{2}$, and it is justified
only when those objects share more than their parameterization. Section
\ref{sec:role-of-parameters} examines what is lost when the parametric
expression is mistaken for the function itself.

\section{The Dollar Principle}

\label{sec:dollar-principle}

The argument so far is largely formal. Before pressing further into
the role of parameters and the geometry of $M$, we pause for an analogy.
The mathematics of the previous section is meant to clarify a situation
that is otherwise prone to confusion, but mathematics alone does not
always convey what is at stake. The following parable does.

Imagine a country in which financial dealings are governed by what
its statutes call \emph{the dollar principle}: two transactions have
the same effect on a person's bank account whenever the dollar amounts
recorded for the two transactions are the same. So formulated, the
principle is innocuous. It is the elementary observation that money
does what money does, and that a deposit of fifty dollars credits
the account by fifty dollars regardless of whether the transaction
was a refund, a paycheck, or a gift. The dollar amount is a numerical
value that points to a quantity of interest---the change in the account---and
within the country the pointing is unambiguous.

Suppose, encouraged by the smooth operation of the dollar principle,
the legislators of this country decide that since dollar amounts work
so well, they should work uniformly. They enact \emph{the strong dollar
principle}: two transactions have the same effect on a bank account
whenever the dollar amounts recorded for the two transactions are
the same, regardless of which country's dollars are at issue. Henceforth,
a transaction of fifty Canadian dollars and a transaction of fifty
US dollars are to be treated as equivalent for all purposes, because
the recorded numerical values agree. A still more ambitious legislator
might propose the further extension that the year of the transaction
be ignored, so that fifty dollars in 1925 and fifty dollars today
are also to be treated as equivalent.

The strong dollar principle is nonsense, and obviously so. A Canadian
dollar and a US dollar are not the same quantity of value; they are
quantities of value measured in different currencies. The numerical
agreement of their recorded amounts is an artifact of a shared symbol,
not a fact about the underlying quantities. The original dollar principle
worked because, within the country, the symbol ``dollar'' picked
out a single thing. Strengthening the principle to range across countries
severs the symbol from the thing it was naming, and the principle
ceases to be about the underlying quantity at all. It becomes a principle
about the recorded representation, mistaken for a principle about
what is recorded.

This is the heart of the matter. Section \ref{sec:role-of-parameters}
develops the analogy formally, treating parameters as playing the
role of units of measurement and showing which features of likelihood-based
inference are coordinate-free.

\section{Role of Parameters}

\label{sec:role-of-parameters}

The likelihood $L^{y}_{M}$ is defined on $M$, and the principal
quantities of inference---the MLE, KL divergence, Hellinger distance---are
functions of distributions, not of parameter values. Why, then, do
parameters appear at all?

The reason is that $M$ generally carries more structure than a set.
Most families of interest are smooth manifolds: the distributions
in $M$ vary smoothly with one another, and quantities of inferential
interest (the gradient of the log likelihood, the curvature of $\ell^{y}_{M}$
at the MLE, the metric structure that defines distance between distributions)
are differential objects on this manifold. Differential calculus on
a manifold requires coordinates. A parameterization $\theta_{M}:M\rightarrow\Theta$
is a choice of coordinate system: it identifies each $m\in M$ with
a point of $\Theta\subseteq\mathbb{R}^{p}$ and thereby allows the
apparatus of multivariable calculus to be used to study $M$.

The parameterization must be a diffeomorphism, so that the smooth
structure of $M$ is faithfully represented by the smooth structure
of $\Theta$. Which diffeomorphism is chosen does not matter. The
sample $y$ is fixed throughout this section: it selects the log-likelihood
function once and for all, and the parameter is the variable. For
each parameterization $\theta_{M}$ there is a corresponding parametric
log likelihood 
\[
\ell^{y}_{\Theta}=\ell^{y}_{M}\circ\theta^{-1}_{M},
\]
and any other smooth parameterization $\xi_{M}$ gives $\ell^{y}_{\Xi}=\ell^{y}_{M}\circ\xi^{-1}_{M}=\ell^{y}_{\Theta}\circ(\theta_{M}\circ\xi^{-1}_{M})$.
The two parametric functions are equivalent descriptions of the same
object $\ell^{y}_{M}$, expressed in different coordinates. The underlying
function on $M$ is what is being described; the parameter is a label.

This is the same situation as units of measurement on a physical quantity.
A length can be expressed in inches or in centimeters; the numerical
values differ, but the length is the same. The unit is a choice of
coordinate on the one-dimensional manifold of lengths, and the rules
of unit conversion are precisely the diffeomorphism that connects
the two coordinate systems. We illustrate next with the binomial family.

\subsection*{Binomial example: log likelihood in two coordinates}

Let $M$ be the binomial family with index $n$. The sample space
is $\mathcal{Y}=\{0,1,\ldots,n\}$ and a typical $m\in M$ has density
\[
f_{m}(y)=\binom{n}{y}p^{y}(1-p)^{n-y}\qquad\text{for some }p\in(0,1).
\]
We consider two common parameterizations: 
\[
\theta_{M}(m)=p,\qquad\xi_{M}(m)=\log\bigl(p/(1-p)\bigr).
\]
Both are diffeomorphisms onto their images, $(0,1)$ and $\mathbb{R}$
respectively. For computation we fix $n=20$ and the observed count
$y=8$.

\begin{figure}[h]
\centering \includegraphics[width=0.95\linewidth]{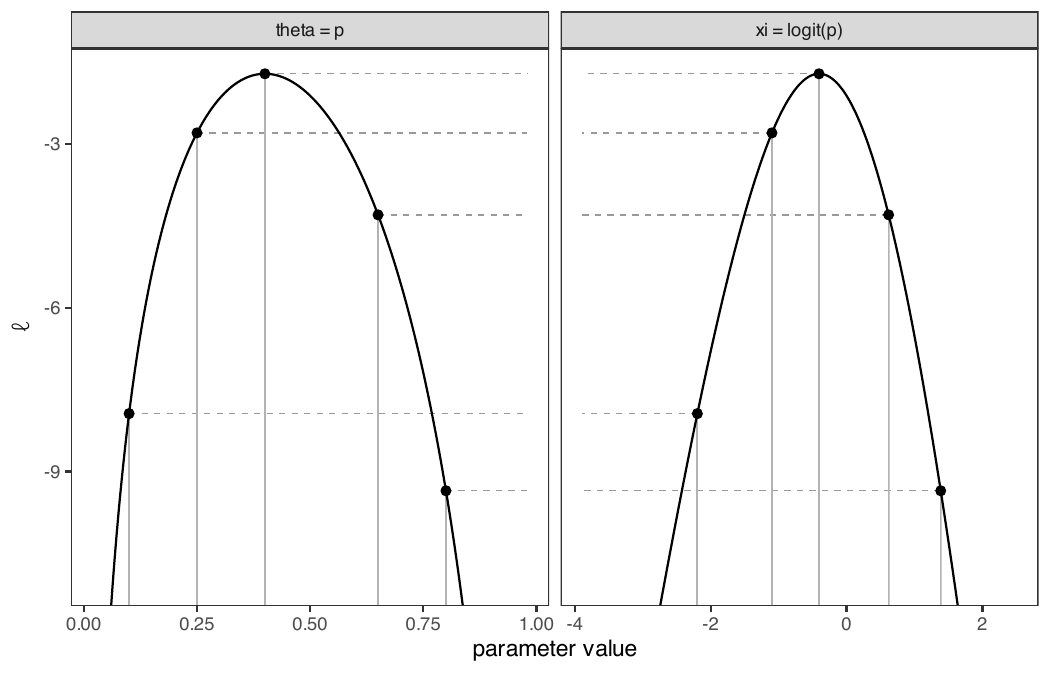}
\caption{Binomial log likelihood plotted against $\theta=p$ (left) and $\xi=\mathrm{logit}(p)$
(right), for $n=20$, $y=8$. Gray segments rise from the floor to
five reference distributions; dashed horizontals connect the two coordinate
pictures of each $m$. The heights at matched segments agree: $\ell^{y}_{M}$
is one function on $M$. The slopes differ, because they are expressed
in different units.}
\label{fig:loglik} 
\end{figure}

Figure \ref{fig:loglik} shows the parametric log likelihoods $\ell^{y}_{\Theta}$
and $\ell^{y}_{\Xi}$ in their respective coordinates. Five reference
distributions are marked to illustrate parameter-invariance, with
each $m$ appearing as a vertical segment in the left panel (at $\theta_{M}(m)$)
and in the right panel (at $\xi_{M}(m)$). The heights of the curve
at corresponding segments agree across the two panels---they record
the value $\ell^{y}_{M}(m)$, which does not depend on the coordinate.
The shapes of the two curves differ, however, and in particular their
slopes differ. The slope of $\ell^{y}_{\Theta}$ measures rate of
change per unit of $\theta$; the slope of $\ell^{y}_{\Xi}$ measures
rate of change per unit of $\xi$. These are different units, so the
slopes are numerically different, but they describe the same underlying
rate of change of $\ell^{y}_{M}$ at $m$.

\subsection*{The score in a coordinate, the score on $M$}

We recall that $\theta_{M}(m)=p$ and $\xi_{M}(m)=\log(p/(1-p))$
are both functions on $M$ of the same underlying probability $p$.
The score is the derivative of the log likelihood with respect to
the parameter; we write it $\partial\ell^{y}_{\Theta}$, where the
symbol $\partial$ stands for the partial derivative with respect
to the parameter identified by the domain subscript on $\ell$. So
$\partial\ell^{y}_{\Theta}$ is shorthand for $\partial\ell^{y}_{\Theta}/\partial\theta$,
and similarly $\partial\ell^{y}_{\Xi}$ for $\partial\ell^{y}_{\Xi}/\partial\xi$;
in each case $\partial$ takes its meaning from the subscript on the
function it acts on. The score $\partial\ell^{y}_{\Theta}:\Theta\rightarrow\mathbb{R}$
is a function on $\Theta$; to evaluate it at a point of $M$ we compose
with $\theta_{M}$. The two parametric scores are 
\[
\partial\ell^{y}_{\Theta}\circ\theta_{M}(m)=\frac{y}{p}-\frac{n-y}{1-p},\qquad\partial\ell^{y}_{\Xi}\circ\xi_{M}(m)=y-np,
\]
and the corresponding Fisher information functions $I_{\Theta}:\Theta\rightarrow\mathbb{R}^{+}$
and $I_{\Xi}:\Xi\rightarrow\mathbb{R}^{+}$ are 
\[
I_{\Theta}\circ\theta_{M}(m)=\frac{n}{p(1-p)},\qquad I_{\Xi}\circ\xi_{M}(m)=np(1-p).
\]
At a reference distribution $m_{\circ}$ with $p_{\circ}=0.5$, the
two scores take different numerical values---specifically, $\partial\ell^{y}_{\Theta}\circ\theta_{M}(m_{\circ})=-8$
and $\partial\ell^{y}_{\Xi}\circ\xi_{M}(m_{\circ})=-2$. The ratio
$\partial\ell^{y}/\sqrt{I}$, however, takes the same value in both
coordinates, where it is understood that the same parameterization
is used for the numerator and the denominator (so the subscript $\Theta$
or $\Xi$ is the same on top and bottom even though we have suppressed
it). This is not a coincidence. Under a change of parameterization
$\xi=\xi(\theta)$, 
\[
\partial\ell^{y}_{\Xi}=\partial\ell^{y}_{\Theta}\,\frac{d\theta}{d\xi},\qquad I_{\Xi}=I_{\Theta}\Bigl(\frac{d\theta}{d\xi}\Bigr)^{2},
\]
(where $\partial$ on the two sides denotes the derivative in the
respective parameter), so the ratio $\partial\ell^{y}/\sqrt{I}$ is
the same value in all parameterizations. It is a function of $m$,
not of the parameter chosen to label $m$, and so it has the same
value at $m$ in every parameterization.

\begin{figure}[h]
\centering \includegraphics[width=0.95\linewidth]{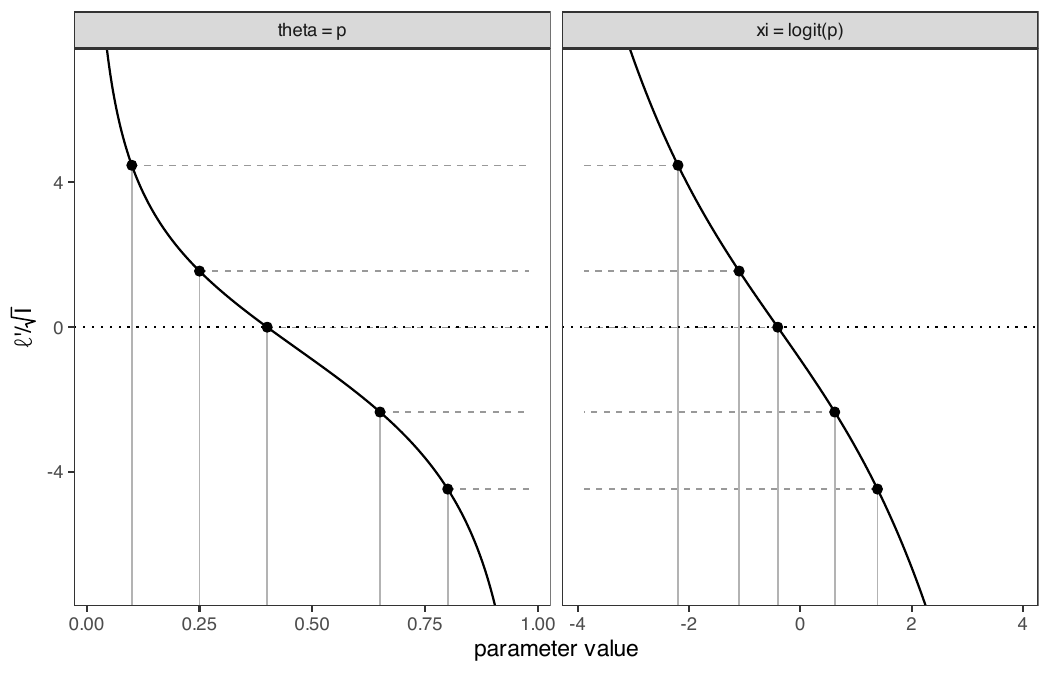}
\caption{Standardized score $\partial\ell^{y}/\sqrt{I}$ for the binomial family
in coordinates $\theta=p$ (left) and $\xi=\mathrm{logit}(p)$ (right).
Heights at the five reference distributions agree across the two panels:
the standardized score is a function on $M$, not on the parameter
space. The curve crosses zero at the MLE, $\hat{m}_{y}$.}
\label{fig:score-std} 
\end{figure}

Figure \ref{fig:score-std} shows the standardized score $\partial\ell^{y}/\sqrt{I}$
in each coordinate. At each of the five reference distributions, the
heights of the curve match across the two panels---in contrast to
the score itself, which would not. The standardized score is a function
on $M$.

\subsection*{Why $\sqrt{I}$ is the conversion factor}

The Fisher information arises as the second-order coefficient of the
Kullback--Leibler divergence. For nearby distributions $m_{\circ},m\in M$
with $\theta_{M}(m_{\circ})=\theta_{\circ}$ and $\theta_{M}(m)=\theta$,
a Taylor expansion of $KL_{\Theta}(\theta_{\circ},\theta)$ at $\theta=\theta_{\circ}$
gives 
\begin{align*}
KL_{\Theta}(\theta_{\circ},\theta) & =-\tfrac{1}{2}E\bigl[\partial^{2}\ell_{\Theta}(\theta_{\circ})\bigr](\theta-\theta_{\circ})^{2}+o\bigl((\theta-\theta_{\circ})^{2}\bigr)\\
 & =\tfrac{1}{2}I_{\Theta}(\theta_{\circ})(\theta-\theta_{\circ})^{2}+o\bigl((\theta-\theta_{\circ})^{2}\bigr),
\end{align*}
the second equality being the definition of the Fisher information.
The KL divergence itself is a function of $(m_{\circ},m)$, so defining
\[
d(m_{\circ},m)=\sqrt{2\,KL_{M}(m_{\circ},m)}
\]
gives a quantity on $M\times M$ that, to leading order, behaves like
a distance: 
\[
d(m_{\circ},m)=\sqrt{I_{\Theta}(\theta_{\circ})}\,|\theta-\theta_{\circ}|+o(|\theta-\theta_{\circ}|).
\]
Substituting into the Taylor expansion of $\ell^{y}_{\Theta}$ around
$\theta_{\circ}$ gives 
\begin{eqnarray*}
\ell^{y}_{M}(m)-\ell^{y}_{M}(m_{\circ}) & = & \frac{\partial\ell^{y}_{\Theta}(\theta_{\circ})}{\sqrt{I_{\Theta}(\theta_{\circ})}}\,d(m_{\circ},m)+o\bigl(d(m_{\circ},m)\bigr)\\
 & = & \frac{\partial\ell^{y}}{\sqrt{I}}(m_{\circ})\,d(m_{\circ},m)+o\bigl(d(m_{\circ},m)\bigr),
\end{eqnarray*}
where $\partial\ell^{y}/\sqrt{I}:M\rightarrow\mathbb{R}$ is well-defined
because it takes the same value at $m$ in every parameterization.
The factor $\sqrt{I_{\Theta}(\theta_{\circ})}$ converts a parameter
increment $|\theta-\theta_{\circ}|$ into a manifold distance $d(m_{\circ},m)$.
Equivalently, $\partial\ell^{y}_{\Theta}$ measures rate of change
per unit of $\theta$; dividing by $\sqrt{I_{\Theta}}$ gives rate
of change per unit of $d$, a quantity intrinsic to $M$.

In differential-geometric language, $I_{\Theta}$ is the coordinate
expression of a Riemannian metric on $M$ (the Fisher information
metric, induced by the KL divergence), and $\sqrt{I_{\Theta}}$ relates
infinitesimal increments in the coordinate $\theta$ to infinitesimal
arc lengths on $M$. The standardized score $\partial\ell^{y}/\sqrt{I}$
is the gradient of $\ell^{y}_{M}$ with respect to this metric---an
object on $M$, with the same value at $m$ in whatever coordinate
one happens to be using.

\subsection*{Negative binomial: same parameter, different family}

The negative binomial family $N$ with fixed number of successes $r$
shares the parameter $p\in(0,1)$ with the binomial. For matched data---$n=20$,
$y=8$ for the binomial; $r=8$, $k=12$ for the negative binomial---the
two likelihoods are proportional as functions of $p$: 
\[
L^{k}_{\mathrm{NB}}(p)=\frac{r}{n}\,L^{y}_{\mathrm{Bin}}(p),
\]
where we use the family label as a subscript to distinguish parametric
quantities derived from the two families on the shared parameter space
$(0,1)$. This is the strong likelihood principle in this example:
viewed as functions of the shared parameter $p$, the two likelihoods
agree up to a positive multiplicative constant, and the SLP asserts
that any inference depending only on this shared parametric form should
treat the two cases as equivalent.

\begin{figure}[h]
\centering \includegraphics[width=0.85\linewidth]{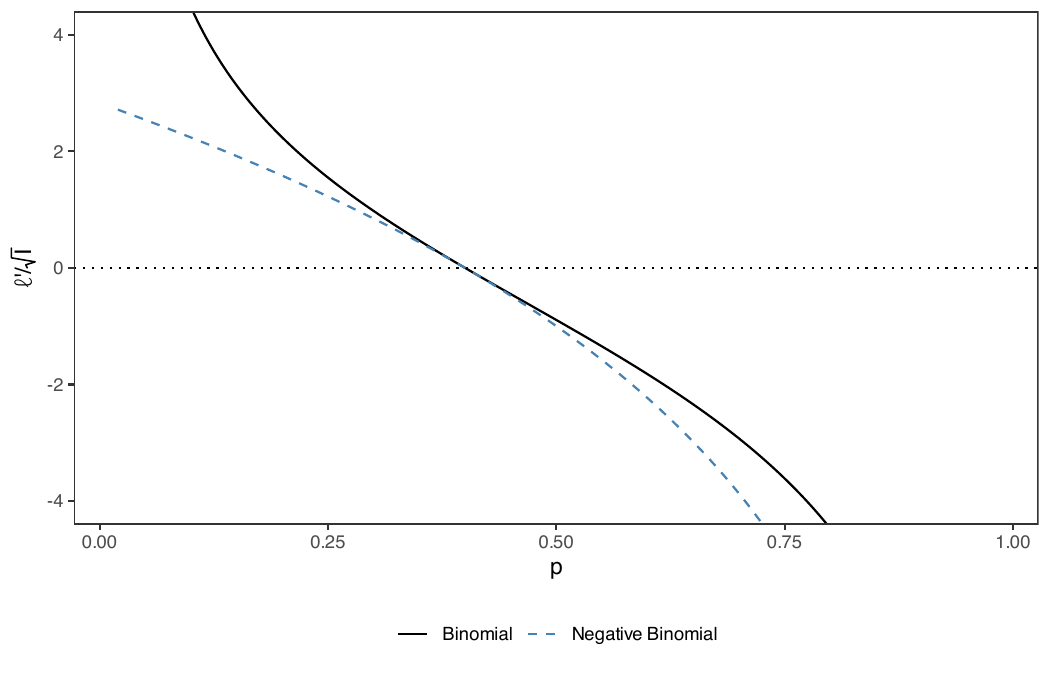}
\caption{Standardized score $\partial\ell^{y}/\sqrt{I}$ for the binomial and
negative binomial families, plotted against the shared parameter $p$,
for matched data. The two parametric log likelihoods agree as functions
of $p$ up to a constant, but the standardized scores are different
functions of $p$ because the Fisher informations differ.}
\label{fig:std-score-compare} 
\end{figure}

The standardized scores, however, do not agree. Taking logarithms
turns the multiplicative constant into an additive one, which differentiation
removes; the score functions are therefore equal as functions of $p$.
The Fisher informations, by contrast, differ: 
\[
I_{\mathrm{Bin}}(p)=\frac{n}{p(1-p)},\qquad I_{\mathrm{NB}}(p)=\frac{r}{p^{2}(1-p)}=\frac{r/n}{p}\,I_{\mathrm{Bin}}(p).
\]
Figure \ref{fig:std-score-compare} plots the two standardized scores
on the common $p$-axis. They are different functions of $p$: at
each value of the shared parameter, the standardized score takes different
values in the two families.

The interpretation is consistent with what was developed in Section
\ref{sec:parameter-free}. The binomial and the negative binomial
are different families, with different sample spaces and different
distributions; the parameter $p$ is the \emph{same symbol} in both,
but it is not the same coordinate system---it is one coordinate system
on $M$ and a different one on $N$. The two manifolds $M$ and $N$
carry different metric structure at corresponding distributions, and
quantities of inferential interest that depend on the metric (the
Fisher information, the standardized score, the KL divergence between
nearby distributions) take different values. The shared symbol $p$
is the illusion of agreement; the manifolds underneath are the reality
of disagreement.

The closed-form KL divergences make this concrete: 
\[
KL_{\mathrm{Bin}}(p_{1},p_{2})=n\!\left[p_{1}\log\frac{p_{1}}{p_{2}}+(1-p_{1})\log\frac{1-p_{1}}{1-p_{2}}\right],
\]
\[
KL_{\mathrm{NB}}(p_{1},p_{2})=r\!\left[\log\frac{p_{1}}{p_{2}}+\frac{1-p_{1}}{p_{1}}\log\frac{1-p_{1}}{1-p_{2}}\right].
\]
The ratio depends only on the base point: 
\[
\frac{KL_{\mathrm{NB}}(p_{1},p_{2})}{KL_{\mathrm{Bin}}(p_{1},p_{2})}=\frac{r/n}{p_{1}},
\]
and the same expression gives the ratio of Fisher informations at
$p_{1}$. The two metrics on the shared parameter set $(0,1)$ are
pointwise multiples of one another, but the multiplier is not constant.
Two distributions that are infinitesimally close in the binomial sense
are not infinitesimally close to the same degree in the negative binomial
sense, even though both are labeled by the same $p$.

This is the formal expression of the diagnosis offered in Section
\ref{sec:dollar-principle}: numerical agreement in a shared coordinate
does not establish equivalence of the underlying objects, because
the objects carry structure that the coordinate alone does not record.

\section{Score Principle and Standardization}

\label{sec:score-principle}

The argument so far has run on geometric considerations: parameters
are coordinates on $M$, the score is a directional derivative whose
numerical value depends on the coordinate, and the standardized score
$\partial\ell^{y}/\sqrt{I}$ is what one obtains by referring this
derivative to the intrinsic metric on $M$. There is a parallel argument
that runs on entirely statistical considerations and arrives at the
same standardization. The likelihood principle is conventionally stated
in terms of the likelihood function, but an equivalent statement uses
the log likelihood: two outcomes are inferentially equivalent if their
log likelihoods agree up to an additive constant. The weak form restricts
the claim to outcomes from the same family and is uncontroversial.
The strong form extends it across families, and the closing argument
of the previous section gave a counterexample even when the families
share their parameter.

The most natural way to compare log likelihoods up to additive constants
is to differentiate them, which removes the constant and produces
the score.  If $m_{\circ}$ and $m$ are nearby distributions in
$M$ and one differentiates $\ell^{y}_{M}(m)-\ell^{y}_{M}(m_{\circ})$
along a path connecting them, the limit is the score. Two log likelihoods
that agree up to a constant therefore yield identical scores, and
the (strong) likelihood principle implies a corresponding (strong)
score principle: that inferences should be the same when the score
functions agree, regardless of whether the scores arise from the same
family or from different families with a shared parameter. As a logical
consequence of the SLP, the score principle inherits whatever force
the SLP itself has. We will argue that the force is illusory.

The argument is straightforward. Two outcomes from the same family
with the same score function describe the same rate of change of $\ell^{y}_{M}$
at the same point, and the weak score principle---the same-family
version---is as benign as the weak likelihood principle. But outcomes
from different families with the same score function are another matter.
The binomial and negative binomial example of Section \ref{sec:role-of-parameters}
made this concrete: the two parametric scores $\partial\ell^{y}_{\Theta}$
agree identically as functions of $p$ (differentiation removed the
constant), yet they describe rates of change with respect to different
metric structures on the shared parameter set. Inference that depends
on this metric---and any inference framed in standard-deviation units
does---takes different values in the two families. The strong score
principle, stated in terms of raw scores, is analogous to the strong
likelihood principle stated in terms of parametric likelihoods: it
confuses agreement of coordinate expressions with agreement of the
underlying objects.

\subsection*{Standardization in statistical practice}

A scalar measurement is rarely comparable across populations in raw
form. Heights in centimeters are compared between populations by standardizing---subtracting
the mean and dividing by the standard deviation---to produce a quantity
that is meaningful on a common scale. The same principle applies to
test scores: two students who both took the SAT can be compared by
their raw scores, but a student who took the SAT and a student who
took the ACT are compared by their standardized scores, because the
two tests put different scales on the underlying construct. The practice
is universal and is taught early; it is so familiar that it is rarely
elevated to a stated principle.

\subsection*{Standardization of the score}

To describe the sampling distribution of the score we allow $y$ to
vary, so it becomes an explicit argument. We write $\ell_{\Theta}(y,\theta)=\log f(y,\theta)$
for the log likelihood as a function of both the data and the parameter;
the function considered in the previous section is its slice at a
fixed observed $y$. The score $\partial\ell_{\Theta}(y,\theta)=\partial\log f(y,\theta)/\partial\theta$,
evaluated at a specific $\theta$, is a random variable in $y$ with
mean zero under $f(\cdot,\theta)$. Standardizing this random variable
means dividing by its standard deviation, so the standardized score
is 
\[
\frac{\partial\ell_{\Theta}(y,\theta)}{\sqrt{V\bigl(\partial\ell_{\Theta}(Y,\theta)\bigr)}}.
\]
The variance in the denominator is, by definition, the Fisher information
$I_{\Theta}(\theta)$, so the standardized score is $\partial\ell_{\Theta}(y,\theta)/\sqrt{I_{\Theta}(\theta)}$.
This is exactly the quantity that emerged from the geometric argument
of Section \ref{sec:role-of-parameters} as the gradient of $\ell^{y}_{M}$
with respect to the Fisher metric. The two derivations---one from
the manifold structure of $M$, one from the standard statistical
practice of standardizing for comparison across populations---arrive
at the same quantity.

\section{Discussion}

The two standardizations of the score---the geometric one, which
divides by the conversion factor $\sqrt{I_{\Theta}}$ from the Fisher
metric, and the statistical one, which divides by the standard deviation
of the score---agree. This agreement is a property of the score in
particular, not of standardization in general: for an arbitrary function
$g$ of $y$ and the parameter, the two standardizations yield different
quantities. \cite{vos2022} shows that the agreement is precisely
what makes the score information-optimal, with the Fisher information
appearing as the bound that the score attains and no other function
does. \cite{vosWu2025} extend this result to the multi-dimensional
case and to inference in the presence of nuisance parameters. Taken
together, these papers supply the positive side of the present argument:
just as we have shown that the strong likelihood principle has no
force beyond the weak likelihood principle, they show that the weak
likelihood principle has positive content, in the form of an information
bound that the score attains and no other function of the data does.
Fisher's claim that the likelihood is the appropriate basis for inference
about $m$ is, on this reading, a theorem.

The case of the dollar principle and the case of the likelihood principle
share a structural feature that has been implicit but not yet stated.
A unit of length and a parameterization of $M$ look like different
sorts of objects---one is a choice of scale on the sample space,
the other a choice of coordinates on a family of distributions---but
both are members of equivalence classes defined by a group action.
Units of length are related by positive affine transformations of
$\mathbb{R}$; smooth parameterizations of $M$ are related by diffeomorphisms.
The underlying object---a length, or a distribution---is what remains
invariant under the group action; the unit or the parameter is one
representative of the class. The strong dollar principle and the strong
likelihood principle both attempt to make claims about the underlying
object using the representative alone, and the same objection blocks
both: the representative is not invariant under the group, and claims
at the level of representatives are not, in general, claims about
the underlying object.

The argument of this paper is not that statisticians ought to abandon
their convictions about SLP, but that the conviction in question is
more accurately described by a different name. What the strong likelihood
principle asserts, when carefully read, is that two outcomes from
distinct families with parametric likelihoods proportional in a shared
parameter should generate the same inference. This is a claim about
parameters, not about likelihoods. If the principle were called the
\emph{parameter principle}, the source of the disagreement would be
more clearly visible: those who accept the principle do so on the
view that parameters, once shared, settle the inferential question;
those who do not accept it deny exactly this. The likelihood, as a
function on $M$, is not the locus of the disagreement; it never was.

\end{document}